\documentstyle[preprint,aps]{revtex}
\begin{document}
\draft
\tightenlines
\title{
\bf{Scaling of fluctuation for  
Directed polymers with random interaction
} }
\author{Sutapa Mukherji$^{a,b}$\cite{eml3}, Somendra M.
Bhattacharjee$^{a,b}$\cite{eml2}, and
 A. Baumg\"artner$^a$\cite{eml1}} 
\address{${}^a$Institut f\"ur Festk\"orperforschung, Forschungzentrum,
J\"ulich, Germany\\ ${}^b$Institute of Physics, Bhubaneswar 751
005, India}
\maketitle
\widetext
\begin{abstract}
Using a finite size scaling form for reunion probability, we show 
numerically the existence of a binding-unbinding transition for Directed 
polymers with random interaction. The cases studied are (A1) two chains
in 1+1 dimensions, (A2) two chains in 2+1 dimensions and (B) three chains 
in 1+1 dimensions. A similar finite size scaling form for fluctuation 
establishes a disorder induced transition with identical exponents for
 cases A2 and B. The length scale exponents in all the three cases are 
in agreement with previous exact renormalization group results.
\end{abstract}
\pacs{0540,6470,3620,0520}
\narrowtext
Disorder is often found to give rise to complicated but rich
phenomena in nature, an excellent example of which is the
existence of the spin glass transition and the spin glass
phase\cite{spingl}.  Needless to say that one has to look beyond
the reality, and study simpler models that retain only the
essential features believed to be responsible for such new
events \cite{spingl}.  Directed polymers{\footnote{ A $d+1$
dimensional directed polymer is a random walk stretched in a
particular direction with fluctuations in the transverse $d$
dimensional space.}} (DP) in random media \cite{kpz,hz,smr} or
with random interactions\cite{smb,sm} provide a fruitful basis
in this regard because of its inherent simplicity.  By virtue of
mappings to nonlinear, noisy surface growth
equations\cite{kpz,lassig}, and other applications, DPs have
become relevant in a broader context.  A crucial result
established for directed polymers in random media\cite{hz} is
the existence of a strong disorder ``spin glass" type phase at
$d=1$, and a lower critical dimension, $d=2$, beyond which a
disorder induced transition exists.  For the random interaction
case there is also a weak to strong disorder
transition\cite{smb,sm,lassig}. Directed polymers are also
simpler than the more complex problem of undirected polymers in
random medium\cite{baumg,bkc}.  Similarly, many heteropolymer
(undirected) problems, especially biopolymers, require random
interaction\cite{gutin}, and DPs with random interaction quite
often serve as a simpler solvable limit\cite{higgs}.

In this paper, we concentrate on the disorder induced transition
in directed polymers with random interaction (RANI), to explore
the scaling behaviour of fluctuation due to quenched disorder.
The essential feature here is the mutual short range interaction
which is random, as might arise when there is a random charge
distribution along the length of the polymer.  A different
version of our model appears in the context of wetting phenomena
in the presence of disorder at $d=1$ \cite{der}.  For the RANI
model, unlike the random medium problem, an exact
renormalization group (RG) can be implemented under certain
conditions\cite{smb,sm} .  We like to verify the length scale
exponents obtained from the perturbative RG.  This verification
is essential for further applicability of RG for the disorder
problem to ensure that there is no nonperturbative effects. For
comparison, one might point out the situation for the random
medium problem.  For the weak to strong disorder transition, RG
\cite{tang,frey} predicts a length scale exponent $2/(d-2)$ for
$d>2$ while numerical calculations \cite{gall,kim} for $d=3$
gave numbers as different as 4.2 and 6.7.  We establish here
that, unlike the random medium case, the RANI model is better
controlled.

For generality let us start with a system of $m$ polymers each
of length $N$ with a random $m$ body ($m$th order multicritical)
interaction.  The model in a continuum formulation is given by
\begin{equation}
{\cal H}_m=\frac{1}{2}\ {\sum_{i=1}^{i=m}
\int_0^N {\left(\frac{\partial {\bf r}_i(z)}{\partial z}\right)}}^2 +
\int_0^N dz\ v_m(1+b(z))\prod_{i=2}^{m} \delta({\bf r}_{ii-1}),
\label{eq:ham} 
\end{equation}
where ${\bf r}_i(z)$ denotes the $d$ dimensional spatial
coordinate of the $i$th polymer at contour length $z$. The first
term denotes the elastic energy part of the Gaussian chains and
the second term denotes the mutual random contact interaction
among the chains. Note that the interaction is always at equal
length with a random part $b(z)$ in the coupling constant. The
randomness is dependent only on the length $z$ at which the
interaction occurs. This is what one expects for random monomer
distribution along the backbone. More general randomness can
arise\cite{smlas} but will not be considered here. In the
following analysis we shall consider a binary distribution for
the disorder with equal probability.

We study this system (actually, a discrete version of this)
numerically by generating the partition function recursively
along the length of the chain.  The cases considered are (A1)
the two chain problem ($m=2$) for $d=1$, (A2) two chains in
$d=2$, and (B) three chains with three body interaction ($m=3$)
for $d=1$.  The reason for choosing the last two cases, as
explained below, is that $d=2$ ($d=1$) is the marginal case for
the two body (three body) interaction problem.  Instead of
analyzing large systems as has been done in Ref. \cite{lassig}
for $m=2$ and $d=1$, we use finite size scaling to calculate the
exponents for various cases.

The pure counterpart of the disordered model, Eq. \ref{eq:ham}
is at present well understood through the renormalization group
analysis\cite{jj,sb} or by solving the Schr\"odinger type
equations for the two body problem\cite{lip}. An exact RG shows
that there is a bound state for any small attraction for $d<
d_c=d^{pure}_m=2/(m-1)$. This binding unbinding transition is
seen over a characteristic length scale $\xi_{\parallel} \sim
\mid v_m\mid^{-\nu}$, with $\nu=2/\mid\epsilon_m\mid$, and
$\epsilon_m=(m-1)(d-d^{pure}_m)$. The length scale
${\xi_{\parallel}}$ is a measure of the average separation
between two contacts along the chain.  For repulsive energy, the
phase is governed by a stable fixed point where the
corresponding scale invariant theory is described by an infinite
strength of the repulsion or a fermion like behavior. For
$d>d^{pure}_m$, a threshold for the strength of the attraction
is required for the bound state while for smaller attraction or
any repulsion, the chains are asymptotically free.  Exactly at
$d=d^{pure}_m$, the system is at its marginal dimension where
the repelling chains are again like free Gaussian chains but
with certain log corrections, and the characteristic length
scale diverges exponentially as the transition point is
approached.

The random system exhibits certain new features. The fact that
the disorder becomes marginally relevant at d=1 for the two body
problem was observed in the context of the disordered wetting
phenomena \cite{der}. The exact renormalization group(RG) analysis
\cite{smb,sm} reveals, besides establishing this marginal
relevance, that there is a disorder induced phase
transition beyond the critical dimension $d_m^{dis}=1/(m-1)$.
The length scale exponent for this transition is found to be
$\xi^{dis}_{\parallel}
\sim (\Delta-\Delta_c^{dis})^{-\nu_r}$ with
$\nu_r=2/(m-1)(d-d_m^{dis})$, $\Delta$ being the variance of disorder. 
The critical value of disorder is $\Delta_c^{dis}$.
Exactly at the critical dimension
$d_m^{dis}$, any small disorder leads to a strong disorder
behaviour over a characteristic length scale
$\xi_{\parallel}^{dis}\sim \exp(\frac{1}{\Delta})$. However, in
none of the above cases it was possible to investigate the
strong disorder limit because of the absence of any perturbative
strong coupling fixed point.  To investigate further, a real
space RG has been performed on the hierarchical lattice that
brought out several new features \cite{sut}.  A dynamic
renormalization group approach also recovered these results and
furthermore showed that there is no anomalous exponent for the
free energy
\cite{lassig}.  This suggests that the effect of disorder
is reflected in the various cumulants of the partition functions
that characterize the phases and phase transitions
\cite{smb,sm}. 

For the pure system, the unbound phase is characterized by the
reunion partition function ${\cal Z}_{R,m}(N)$ which is the sum
of the Boltzmann weights of all paths corresponding to the
meeting of $m$ chains at any point in (transverse) space at
length $N$, the other end being tied together at the origin.
Similarly, one can define a survival partition function
$Z_{S,m}(N)$ for the chains tied at the origin at one end but
located anywhere at the other end.  The ratio of the two defines
the probability of reunion.  It has been shown that the reunion
and the survival partition functions, and hence the probability,
acquire anomalous scaling at the stable fixed point at $d=1$
\cite{fish,reuni}. Detailed RG analysis yields
the d=1 exact (vicious walker) result that ${\cal Z}_{R,2} (N)
\sim N^{-3/2}$, and predicts that for 
marginal cases $d=d_m^{pure}$, the $m$ chain reunion with $m$
body interaction has the form ${\cal Z}_{R,m} (N) \sim N^{-1}
(\ln N)^{-2}$ \cite{reuni,gutt}.  These cases include the two
chain problem at $d=2$ and three chain at $d=1$.  The identical
log correction, originating from the marginality of the
interaction at the appropriate dimensions, is an example of
"Grand Universality" for the directed polymer multicritical
points\cite{sb}. This feature is preserved in the random case
also.

For the pure versions of cases A1, A2 and B, the unbinding
critical point is at $v_m=0$, for which the chains are free.
Therefore, around the transition, one can write a finite size
scaling form
\begin{equation}
N^{(m-1)d/2}{\cal Z}_{R,m}(N) = {\cal F}({\sf
x}),\label{eq:scal}
\end{equation}
where ${\sf x}=\mid v_m\mid N^{1/\nu}$ for two chains in one
dimension and ${\sf x}=\mid v_m\mid \ln N$ for $m=d=2$ and
$m=3,d=1$. A similar scaling form can also be written for the
probabliy of reunion. A typical case is shown in Fig 1.

The partition function for the chains is generated recursively
in their length. For the two chain problem, since the randomness
is only along the $z$ direction, only the relative chain is
considered.  The center of mass is a free Gaussian chain. This
transformation is not possible for three chains and we consider
the full partition function.

For two chains with pair interaction, the recursion relation is
given by
\begin{mathletters}
\begin{equation}
Z({\bf r},t+1) = e^{-\eta} \ \big[\sum_{i=1}^{d} \{Z({\bf r} +
{\hat e_i},t)+Z({\bf r} - {\hat e_i},t) \}+ p Z({\bf r},t)\big
]/(2 d+p),
\label{eq:twoch}
\end{equation}
with ${\hat e_i}$ as unit vectors in the $i$th direction, and
${\bf r}$ denoting the position vector for the end point.  A
weight factor $p$ has been put in for extra weightage of the
relative chain.  We have set $p$ equal to 4.  The random energy
$\eta = v_m\pm \Delta$, with equal probability, whenever ${\bf
r}$ is zero. In this relative coordinate, ${\cal
Z}_{R,2}(N)=Z({\bf 0},N)$, and $Z_{S,2}(N) = \sum_{\bf r} Z({\bf
r},N)$.  For three chains in $d=1$,
\begin{equation}
Z(x,y,z,t+1)=e^{-\eta}\
\left [\sum_{i,j,k=\pm 1} Z(x+i,y+j,z+k,t)\right]/8,
\label{eq:threech}
\end{equation} 	             
\end{mathletters}
where the Boltzmann factor for the randomness contributes only
for $x=y=z$, i.e., only when three chains meet together. All
chains start at origin at $t=0$.  For the three chain case,
${\cal Z}_{R,3}(N)= \sum_{x} Z(x,x,x,N)$, and
$Z_{S,3}(N)=\sum_{x,y,z} Z(x,y,z,N)$. 

The randomness reduces the strength of the interaction. The
effective interaction ${\overline v_m}$, determined by
\begin{equation}
\exp (-\bar v_m)=[\exp (-v_m-\Delta)+\exp(-v_m+\Delta)]/2,
\end{equation}
controls the binding - unbinding transition.  Since for all the
three cases the pure binding-unbinding transition takes place at
zero energy (i.e. Boltzmann factor=1), the transition point for
a fixed $v_m$ is given by
\begin{equation}
v_m=\ln \cosh \Delta_c.\label{eq:delc}
\end{equation}
As a result a bound state can form even for $v_m>0$.  In other
words, a thermal unbinding is possible because randomness
produces attractive pockets.  The scaling of Eq. \ref{eq:scal}
can be rewritten as $N {\langle {\cal Z}_{R,m}(N)\rangle} =
{\cal F}({\sf x})$, where, for cases A2 and B, ${\sf x} =
(\Delta-\Delta_c) \ln N$ for a given $v_m$ with $\Delta_c$
satisfying Eq. \ref{eq:delc}.

We computed the average reunion and survival partition
functions, and also the average probability of reunion where the
probability for a given realization is defined as $P(N) = {\cal
Z}_{R,m}(N)/Z_{S,m}(N)$.  The main distinction is that while the
first two quantities may be considered as annealed averaging,
the last one is strictly a quenched averaged quantity.  For a
given realization, the recursion relation gives exact values
(upto machine precision) of the partitions functions. An
averaging over the disorder is done by considering around 500
samples. The lattice size is chosen such that the average
transverse size of a polymer ($\sim N^{1/2}$) does not exceed it
(to avoid boundary effects).

For a fixed strength of $v_2$, we plot, in Fig. 1, the data for
different $\Delta$, with the scaled variables $\langle{\cal
Z}_{R,m}(N)\rangle N$ as $(\Delta-\Delta_c)\ln N$.  This plot
shows a nice data collapse on the pure curve with $\Delta_c$
chosen as per Eq. \ref{eq:delc} for two different values of
$v_2$.  It verifies the exponential divergence of the length
scale near the transition. We have also checked similar data
collapse for the other cases and various values of $v_m$. The
main inference we draw from this is that for the average chains
in presence of disorder, the nature of the transition remains
the same as the pure system except for a shift in the critical
value.  This also serves as a check on the simulation.

With the above knowledge, we consider a particular value
$\Delta$ and force the system to be strictly at the transition
point of unbinding by choosing appropriate $v_m$.  The specific
quantity we calculate is the fluctuation in the reunion
partition function or the fluctuation in the probability of
reunion.  The raw data for the cumulant ${\langle P^2\rangle}_c$
have been plotted with $N$ for various $\Delta$ with
corresponding $v_m$ determined by Eq. \ref{eq:delc}.  In case
there is a disorder induced transition we expect the following
scaling to hold good close to this critical point
$\Delta_c^{dis}$,
\begin{equation}
\langle P^2\rangle_c=N^{-\phi} {\cal
G}((\Delta-\Delta_c^{dis})N^{1/\nu_r}),
\end{equation}
where, according to the RG prediction of the divergence of the
length scale near the transition point, $\nu_r=1$ for Cases A2
and B.  If we believe in Grand Universality \cite{smb}, the
exponent $\phi$ should also be the same for these two cases.
The plot with ${{\langle P^2\rangle_c}} N^{\phi}$ vs $ N
(\Delta-\Delta_c^{dis})$ indeed shows a data collapse for a
small value of $\Delta_c^{dis}<.05$ and $\phi=2$. The two cases,
A2 and B, for which the pure interaction is marginal, are shown
in Fig 2 and 3.  Numerical accuracy forbids simulations for very
small values of $\Delta$ and therefore the weak disorder phase
cannot be probed in our simulation.

Fig 4 shows the data collapse for the case of two chains at
$d=1$. Since the disorder is expected to be always marginally
relevant, the scaled variables for data collapse, from
RG\cite{smb,sm}, would be $\Delta \log N$.  We obtain a data
collapse for $\phi=.57\pm.03$.  This verifies the exponential
divergence of the length scale for the situation where the
disorder is marginal.

To conclude, we have shown that randomly interacting directed
polymers undergo a binding-unbinding transition and, at the
critical point, the fluctuation in the reunion partition
function or the probability of reunion show scaling behaviour
with length scale exponents that agree with RG prediction. A
finite size scaling form, verified for the pure case, has been
used to study the unbinding transition..  We also obtained the
exponents that describe the fluctuation of the second cumulant.
This exponent $\phi$ is the same for two chains in (2+1)
dimensions and three chains in (1+1) dimensions for both of
which the pure interaction is marginal.

 SMB and SM acknowledge the hospitality of IFF. Their visit 
was supported by the Indo-German project Number PHY-25/1. The 
computations were done at Julich on IBM and DEC alpha computers.

\begin{figure}
\caption{Data collapse for pure and random cases, two chains in
2+1 dimensions. For the pure case, the $y-$axis is $N {\cal Z}_{R,2}(N)$
while for the random case it is $N \langle {\cal Z}_{R,2}(N)\rangle$.  The
$x$ axis is $\mid v_2\mid \log N$ for the pure case but $(
\Delta-\Delta_c) \log N$ for the random case. The legend shows the values
of $v_2$ for the random case.}
\end{figure}
\begin{figure}
\caption{Data collapse for fluctuation of the reunion partition
function. This is for two chains in 2+1 dimensions.  The
exponent $\phi=2$ and $\nu_r = 1$.  The
inset shows the fluctuations for various $\Delta$ and length. }
\end{figure}
\begin{figure}
\caption{Data collapse for reunion probability for three chains
in 1+1 dimensions.The exponents are $\phi=2$ and $\nu_r=1$. The
inset shows the variance for various $\Delta$ and length.}
\end{figure}
\begin{figure}
\caption{Data collapse for fluctuation of the reunion probability
for two chains in 1+1 dimensions.  $\phi=.57\pm .03$. The inset
shows the variance as in Figs 2 and 3.}
\end{figure}
\end{document}